\documentclass[english]{extarticle}
\usepackage{lmodern,ulem}

\usepackage[T1]{fontenc}
\usepackage[latin9]{inputenc}
\usepackage{geometry}
\usepackage{babel}
\usepackage{verbatim,color}
\usepackage{amsmath}
\usepackage{amssymb}
\usepackage{graphicx}
\usepackage[unicode=true,
 bookmarks=false,
 breaklinks=false,pdfborder={0 0 1},backref=section,colorlinks=false]
 {hyperref}
\usepackage{soul} 
\usepackage{xcolor}

\makeatletter

\usepackage{babel}

\makeatother

\begin{document}
\title{How general is the Jensen--Varadhan large deviation functional for 1D
conservation laws?}
\author{Julien Barr\'e$^1$, Ouassim Feliachi$^{1,2}$}

\maketitle
\begin{center}

$^1$ Institut Denis Poisson, Universit\'e d'Orl\'eans, Universit\'e de Tours and CNRS, France.

$^2$ Laboratoire de M\'et\'eorologie Dynamique/IPSL, \'Ecole Normale Sup\'erieure, PSL Research University, Sorbonne Universit\'e,
\'Ecole Polytechnique, IP Paris, CNRS, France.

\end{center}
Corresponding author's email address: julien.barre@univ-orleans.fr

\begin{abstract}
Starting from a microscopic particle model whose hydrodynamic limit under hyperbolic space-time scaling is a 1D conservation law,
we derive the large deviation rate function encoding the probability to observe a density profile which is a non entropic shock, and compare this large deviation rate function with the classical Jensen-Varadhan functional, 
valid for the totally asymmetric exclusion process and the weakly asymmetric exclusion process in the strong asymmetry limit.  
We find that these two functionals have no reason to coincide, and in this sense Jensen-Varadhan functional is not universal. However, they do coincide in a small Mach number limit, suggesting that universality is restored in this regime. We then compute the leading order correction to the Jensen-Varadhan functional.
\end{abstract}

\section{Introduction}
Large deviation principles (LDP) around a hydrodynamic limit are
well studied when the deterministic description is a diffusion equation: they constitute the basis of the macroscopic fluctuation theory (MFT), by now standard to describe fluctuations in out of equilibrium diffusive systems \cite{MFT,MFT2}. Such LDPs have been much less studied however when the system is considered in the ballistic, or "Euler", scaling, and the deterministic equation is then often a hyperbolic conservation law. 

We may identify two approaches for this question. First, the Ballistic Macroscopic Fluctuation Theory (BMFT) has been introduced recently \cite{Doyon19,Doyon22a,Doyon22b} precisely to adapt some ideas and methods of the MFT to the hyperbolic case (see also \cite{SpohnBook} chapter 4.3 for an earlier discussion).  
The crucial remark is that, under the ballistic scaling, no noise term appears in the hydrodynamic equation, which is in this case usually a hyperbolic conservation law. Euler equations of hydrodynamics is the paradigmatic example. As a consequence, initial fluctuations are just transported by the hydrodynamic equation.
BMFT successfully describes correlations of macroscopic variables in various situations (see for instance \cite{Doyon22b}), including in particular the hydrodynamic limit of integrable systems, but is currently restricted to regular evolutions: this is a serious drawback since shocks are a generic feature of (non integrable) hyperbolic conservation laws. 

Hence a second approach to generalize MFT for hyperbolic system focuses on the shocks and other non regular evolutions: this is the one we follow in this article.
This line of research was initiated by Jensen and Varadhan, who rigorously computed the dynamical rate function
for the Totally Asymmetric Simple Exclusion Process (TASEP) \cite{J,JV}. To understand their result, we need to recall that for a given initial condition, a hyperbolic conservation law typically has many weak solutions, among which one, called the entropic solution, is singled out and corresponds to the deterministic evolution; these classical facts are stated in more details in section \ref{sec:conservation}. Jensen and Varadhan show that in the limit of a large number of sites $N$ 
the occupation measure $\rho$ for the TASEP follows a LDP with speed $N$ (number of sites) and rate
function $I$, such that $I[\rho]=+\infty$ if $\rho$ is not a weak
solution of the conservation law; $0<I[\rho]<+\infty$ if $\rho$
is a non entropic weak solution of the conservation law; $I[\rho]=0$
if $\rho$ is an entropic solution. They provide an explicit formula
for $I[\rho]$: the rate function measures in a precise sense "how much non entropic" a
weak solution is. The full proof of the large deviation principle, which was not complete in \cite{J,JV}, is given in \cite{QT}.
While these works are done for the TASEP, \cite{BodineauDerrida2006} starts from the LDP for the Weakly Asymmetric Exclusion Process (WASEP) (see \cite{EnaudDerrida2004,Bertini09}), which has a diffusive deterministic limit. The authors of \cite{BodineauDerrida2006} show that the Jensen Varadhan LDP is then recovered in the strong asymmetry (or equivalently small diffusion) limit (see also \cite{BertiniQP09} for a rigorous study of the quasi-potential in this setting). This is a surprising result, since the WASEP LDP is obtained using a diffusive scaling. 

These results in \cite{BodineauDerrida2006} have been generalized and put on a more rigorous footing by Mariani and coworkers \cite{Mariani10,BBMN}: they start from a diffusion-like
LDP, with a rate function similar to the one for WASEP, and show that this rate function, appropriately rescaled, converges in the small diffusion limit to a generalized Jensen-Varadhan functional, which then describes the probabilistic weight of non entropic shocks. 
This convergence of functionals is in the sense of $\Gamma$-convergence. 

From these works, a general picture of a large deviation theory for hyperbolic conservation laws, including shocks, seems to emerge, based on Jensen-Varadhan functional and its generalizations.   
However, on one hand Jensen, Varadhan, Bodineau and Derrida results rely on specific microscopic models; on the other hand, Mariani et al. approach starts from a fairly general diffusive LDP, but which connection to a  
microscopic dynamics a priori requires a diffusive scaling. This raises the natural question: how general actually is Jensen-Varadhan functional? Or: can we exhibit a system whose hydrodynamic limit is a hyperbolic conservation law and whose fluctuations are not described by a Jensen-Varadhan like functional? Furthermore, a hyperbolic conservation law complemented with a small diffusion can be seen as a simplified caricature of compressible Navier-Stokes equations. Thus one more long-term motivation is to understand the large deviation structure of the compressible Navier-Stokes equation \cite{These_Ouassim}. 

\begin{figure}
\centering{\includegraphics[clip, trim=5cm 17.7cm 3cm 5cm]{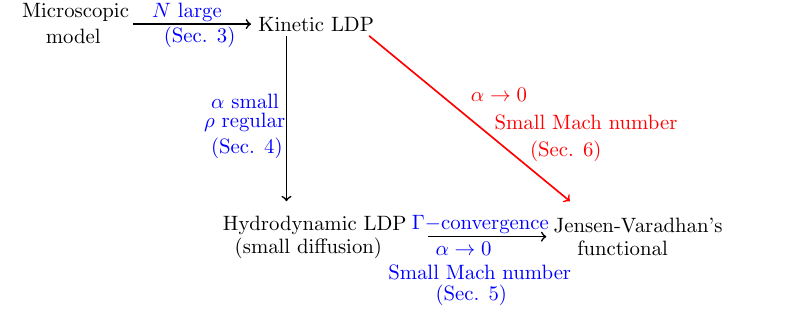}}
 \caption{\label{fig:intro} Summary of this article. Section \ref{sec:model} presents the microscopic model and the large deviation principle (LDP) at the kinetic level, starting point of the following; these results do not require a small $\alpha$.
 Section \ref{sec:-convergence-neglecting-shocks}  
 derives an LDP for the density $\rho$ for small but non zero Knudsen number $\alpha$ (only the formally leading order is kept); the deterministic dynamics is then a conservation law with a small diffusion.
 This section requires that the density profiles are regular enough.
Section \ref{sec:JV} then derives the Jensen-Varadhan functional for $\alpha \to 0$; this is essentially an informal presentation of 
the results of \cite{Mariani10,BBMN}. It yields Jensen-Varadhan's functional, but is not consistent with Sec. \ref{sec:-convergence-neglecting-shocks} without a small Mach number hypothesis. The direct computations of Section
\ref{sec:direct} \textit{do not} yield Jensen-Varadhan's functional, unless the small Mach number hypothesis is made.}
\end{figure}

In order to address the above question, our strategy is to analyze
in details a simple microscopic model, very different from the TASEP, which can be described mesoscopically
by a kinetic equation and a kinetic large deviation principle. We introduce the Knudsen number $\alpha=\ell/L$ where $\ell, L$ are respectively microscopic (mean free path) and macroscopic lengthscales. We
then analyze in the small Knudsen number limit the probabilistic cost of a non entropic shock.
This can be done in two ways: either through a Chapman-Enskog like expansion at the level of large deviation functions leading to a hydrodynamic LDP at small but finite $\alpha$ similar to the starting point of \cite{Mariani10}, followed by a $\Gamma$-convergence at $\alpha \to 0$ computation;
or directly from the kinetic LDP, trying to take advantage of the smallness of $\alpha$. The first route is similar to taking the strong asymmetry limit in the WASEP LDP, while the second one is akin to directly tackling the TASEP, as done by Jensen and Varadhan. These two routes are schematized on Fig.\ref{fig:intro}.

Our main results are: i) The first route, which always yields Jensen-Varadhan's functional, is not consistent in general: indeed the first step (Chapman-Enskog expansion) requires a regular enough density profile, while the second step ($\Gamma-$ convergence for small Knudsen number $\alpha$) concentrates the probability on singular profiles; ii) without further hypotheses, the second route does not yield Jensen-Varadhan's functional; iii) we identify a regime where indeed theses two routes coincides. In other words, Jensen-Varadhan's large deviation functional seems to be valid for our model only in a specific regime, corresponding to a small enough Mach number; iv) we compute the first correction to Jensen-Varadhan functional in the small Mach number regime.

Finally, we notice that taking first a diffusive scaling, as done for the WASEP in \cite{BodineauDerrida2006}, is in a sense a small Mach number regime: the sound velocity is sent to infinity. This is consistent with the fact that Jensen-Varadhan functional is recovered in the strong asymmetry (or equivalently small diffusion) limit. However we have no physical intuition on why Jensen-Varadhan functional is exactly valid for the TASEP. We point out that, although we hope they provide more general ideas, our results rely on a specific microscopic model, and should be investigated in other settings. We also remark that our starting point is a large $N$ large deviation principle, so that the $\alpha \to 0$ limit is taken after the $N\to \infty$ limit, see Fig.\ref{fig:intro}. One may wonder if joint limits $N\to \infty, \alpha \to 0$ could be taken, possibly even restoring Jensen-Varadhan's functional in some specific regime.

\section{Conservation laws, shocks, antishocks, small viscosity and Jensen--Varadhan theory}
\label{sec:conservation}

In this section, we recall basic facts on hyperbolic conservation laws, and present the Jensen-Varadhan large deviation rate function.
We consider a $1D$ hyperbolic conservation law 
\begin{equation}
\partial_{t}\rho+\partial_{x}\left(a\left(\rho\right)\right)=0,\label{eq:invisci}
\end{equation}
where we shall often assume $a:\rho\mapsto a(\rho)$ to be convex (or concave). We call $\rho(x,t)$
a weak solution of (\ref{eq:invisci}) if for all regular enough functions
$\varphi(x,t)$ with compact support in $\mathbb{R}\times\left[0,T\right]$,
the following identity holds 
\[
\int_{\mathbb{R}}\text{d}x\int_{0}^{\infty}\text{d}t\,\left(\partial_{t}\rho+\partial_{x}\left(a\left(\rho\right)\right)\right)\varphi=0.
\]
Note that if $\rho$ is not regular enough, we make sense of this
previous identity with partial integration: 
\[
\int_{\mathbb{R}}\text{d}x\int_{0}^{\infty}\text{d}t\,\left(\rho\partial_{t}\varphi+a\left(\rho\right)\partial_{x}\varphi\right)-\int_{\mathbb{R}}\text{d}x\,\rho\left(x,0\right)\varphi\left(x,0\right)=0.
\]
With an appropriate choice of initial data, (\ref{eq:invisci}) is
known to exhibit discontinuous weak solutions called shocks, the simplest of which are traveling steps: 
\[
\rho(x,t)=\begin{cases}
\rho_{L} & \text{if}\,x\leq v_{s}t,\\
\rho_{R} & \text{if}\,x>v_{s}t,
\end{cases}
\]
where the shock velocity $v_s$ is given by the Rankine-Hugoniot relation: $v_{s}=\left(a\left(\rho_{L}\right)-a\left(\rho_{R}\right)\right)/\left(\rho_{L}-\rho_{R}\right).$
However, it is well known that uniqueness fails for the initial value problem associated to weak solutions of (\ref{eq:invisci}), which indicates that (\ref{eq:invisci}) misses some relevant physics.

\begin{figure}[ht]
\centering{\includegraphics[width=0.5\textwidth]{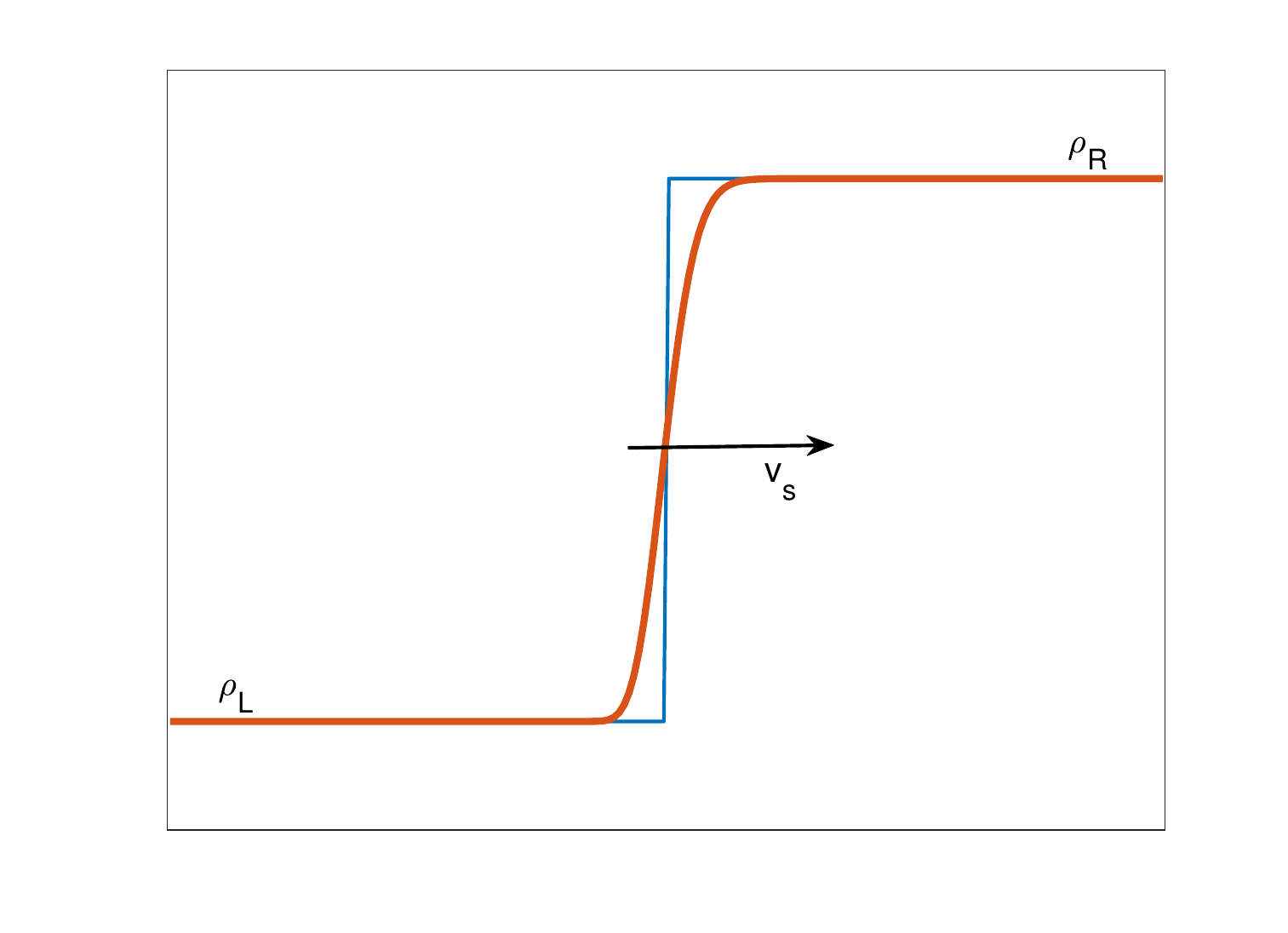}}
\caption{\label{fig:shock} Sketch of an antishock $\rho^A$ for convex flux $a$, traveling at velocity $v_s$ (thin step-like line), and an approximating regularized profile $\rho^\alpha$ (thick smooth line). For a concave flux, an antishock would have $\rho_L>\rho_R$.  }
\end{figure}


If we look at (\ref{eq:invisci}) as the hydrodynamic limit of a
particles system, $\rho$ is a density and $a(\rho)$ is a particle
flux. $a'(\rho)$ is the macroscopic velocity at density $\rho$. $a$ convex
implies that $a'(\rho_{L})>a'(\rho_{R})$ if and only if $\rho_{L}>\rho_{R}$.
Then, this step-like solution is a physical solution if and only if $\rho_{L}>\rho_{R}$:
the particles on the left "catch up" on the particles on the right.
This is why we designate such solutions as \textit{shocks}, whereas
step solutions with $\rho_{L}<\rho_{R}$, a priori unphysical, are called \textit{antishocks}, see Fig.\ref{fig:shock} for an illustration.
\footnote{The situation is simply reversed if $a$ is concave; it is more complicated when $a$ is neither convex nor
concave.}.

A way to discriminate shocks from antishocks is to introduce a small
diffusion term, with $D>0$: 
\begin{equation}
\partial_{t}\rho+\partial_{x}\left(a\left(\rho\right)\right)=\alpha\partial_{x}\left(D\left(\rho\right)\partial_{x}\rho\right).\label{eq:viscous}
\end{equation}
In the small $\alpha$ limit, we formally retrieve the conservation law. When
$\alpha$ is non zero, regularized shocks are still solutions to (\ref{eq:viscous})
but regularized antishocks are not anymore. 

Now, when thinking about
the conservation law as the hydrodynamic limit of a particle system, it makes sense to take into account finite size effects by adding a small conservative noise to (\ref{eq:viscous})\footnote{We use here for the mobility $\sigma$ the same  notation as \cite{BBMN}; the definition of $D$ differs by a $\frac12$ factor.}:
\begin{equation}
\partial_{t}\rho+\partial_{x}\left(a\left(\rho\right)\right)=\alpha\partial_{x}\left(D\left(\rho\right)\partial_{x}\rho\right)+\sqrt{\alpha^2 \varepsilon}\partial_{x}\left(\sqrt{\sigma\left(\rho\right)}\eta\right),\label{eq:SPDE}
\end{equation}
where $\varepsilon$ is a small parameter related to the number $N$ of particles, and $\eta$ is Gaussian and fully characterized by 
\[
\mathbb{E}\left(\eta\left(x,t\right)\eta\left(x',t'\right)\right)=\delta\left(x-x'\right)\delta\left(t-t'\right).
\]
Notice the $\sqrt{\alpha^2}$ factor in the noise term: we could have absorbed it in the currently vaguely defined $\varepsilon$, but we prefer to single it out since it will appear in this form later, and to recall that this noise term vanishes in the $\alpha\to 0$ limit, within Euler scaling: all noise then formally disappear and one recovers the "bare" conservation law \eqref{eq:invisci}. 
While nonlinear SPDEs like \eqref{eq:SPDE} are common in the physics literature, a regularization is necessary to ensure they have a precise mathematical meaning. One may also consider, and this is the point of view we shall adopt, that \eqref{eq:SPDE} is a formal way to write a large deviation principle describing the finite size fluctuations
of the density around the deterministic solution of Eq.\eqref{eq:viscous}. To be precise, we shall consider that \eqref{eq:SPDE} defines a probability measure $\mathbb{P}^{\varepsilon,\alpha}$ on the paths $\rho(x,t)$ over the time interval $[0,T]$, such that $\mathbb{P}^{\varepsilon,\alpha}$ satisfies the large deviation principle
\begin{equation}
\mathbb{P}^{\varepsilon,\alpha}(\rho(\cdot,\cdot)) \underset{\varepsilon \to 0}{\asymp} {\rm exp}\left(\varepsilon^{-1} \alpha^{-2} J_{[0,T]}^{\alpha}[\rho] \right),
\label{eq:ldp_viscous}
\end{equation}
where $\asymp$ stands for the asymptotic equivalence of both sides of the expression in the $\varepsilon \to 0$ limit. The large deviation speed is $\varepsilon^{-1}$ and the large deviation rate function is $\alpha^{-2}J_{[0,T]}^{\alpha}$, with
\begin{equation}
J_{[0,T]}^{\alpha}[\rho]=\frac{1}{2}\int_{0}^{T} ||\partial_{t}\rho+\partial_{x}\big(a(\rho)\big)-\alpha\partial_{x}\big(D(\rho)\partial_{x}\rho\big)||^2_{-1,\sigma(\rho)} \,dt.
\label{eq:ldf_diff}
\end{equation}
Here $||\cdot||_{-1,\sigma}$ is given by (this is the $H^{-1}$ negative Sobolev norm weighted by $\sigma$): 
\begin{equation}
||g ||^2_{-1,\sigma} = \inf_{h} \left\{\int_{\mathbb{R}} \frac{h^2}{\sigma}~:~h'=g\right\}.
\end{equation}
Rate functions similar to \eqref{eq:ldf_diff} are classical for diffusive systems: they underly the MFT \cite{MFT}, and (without the transport term) describe for instance the large deviations around MacKean-Vlasov limit \cite{DawsonGartner87}. Although \eqref{eq:ldp_viscous} is written in an informal manner, it can be given a precise mathematical sense. 
Clearly, $J_{[0,T]}^{\alpha}[\rho]=0$ if and only if $\rho$ is a solution to the PDE \eqref{eq:viscous}. 

The fluctuating hydrodynamics PDE \eqref {eq:SPDE} or the LDP \eqref{eq:ldf_diff} allow to answer the question: what is the probability to observe a given spatio-temporal profile $\rho(x,t)$, which is not necessarily a solution of (\ref{eq:viscous})?
In the small noise/viscosity limit $\alpha\rightarrow0$, it turns out that weak solutions, shocks and antishocks, are overwhelmingly more probable than other profiles, and that, among weak solutions, shocks are overwhelmingly more probable than antishocks. The large deviation principle encoding the probability of antishocks has large deviation speed $(\varepsilon \alpha)^{-1}$, and its rate function is Jensen-Varadhan functional. Consider a shock or antishock  $\rho^A$  with a step profile between density $\rho_L$ on the left and $\rho_R$
on the right, propagating at velocity $v_s$, given by Rankine-Hugoniot condition (see Fig.\ref{fig:shock}). Jensen-Varadhan functional, over a time interval of length $T$, is then: 
\begin{equation}
I_{JV}\left[\rho^{A}\right]=T \int_{{\rm min}(\rho_{L},\rho_R)}^{{\rm max}(\rho_L,\rho_{R})}\text{d}u\,\frac{2 D(u)}{\sigma\left(u\right)}\big[({\rm sgn}(\rho_R-\rho_L)\big(a\left(\rho_{L}\right)-a(u)-v_{s}(\rho_{L}-u)\big)\big]_+.\label{eq:JVLDF0}
\end{equation}
where the $_+$ subscript denotes the positive part  ($x_+=\max(0,x)$). 
Notice that the quantity $v_{s}\rho_{L}-a\left(\rho_{L}\right)-v_{s}u+a(u)$ is negative for a shock profile, hence the rate function vanishes.
This rate function measures the "entropy production", with the entropy
function $s$, such that $s''(u)=D(u)/\sigma(u)$.
Deriving the rate function \eqref{eq:JVLDF0} from the rate function \eqref{eq:ldf_diff} is precisely the question addressed in \cite{Mariani10,BBMN},
and then \cite{Mariani_greek,BBC} in more than one dimension. We recall this computation for completeness in Section \ref{sec:JV} and Appendix \ref{sec:appendixB}. 

\section{Microscopic and kinetic model}

\label{sec:model} In this section, we introduce a simple 
toy-model of interacting particles admitting a $1D$ conservation law as a hydrodynamic
limit, and which large deviation structure is formally tractable. It can be seen as a one-dimensional run and tumble model with density dependent tumbling rates.
Run and tumble dynamics is widely used in biology and biophysics to describe the behavior of bacteria (see for instance \cite{Berg72,Polin09}); we only use it here as a simple tractable model. 

\paragraph{Microscopic model.}

We consider $N$ particles in a domain of size $L$ with positions $\left(x_{n}\right)_{1\leq n\leq N}\in [0,L]^{N}$; boundary conditions will play little role in the following, and be taken as periodic. 
Particles travel with velocity $v$ or $-v$, and switch velocity at a rate depending on the local concentration
$\bar{\rho}_{N,R}(x,t)$: 
\[
\bar{\rho}_{N,R}(x,t) = \frac{L}{2NR}\sum_{i=1}^N \chi\left(\frac{x_i-x}{R}\right),
\]
$\chi$ is the characteristic function of $[-1,1]$.
The particles velocities evolve
according to the Markov jump process:
\begin{equation}
\begin{array}{ccc}
+v & \stackrel{\lambda g_{-}\left(\bar{\rho}_{N,R}\right)}{\longrightarrow} & -v,\\
-v & \stackrel{\lambda g_{+}\left(\bar{\rho}_{N,R}\right)}{\longrightarrow} & +v,
\end{array}\label{eq:particle_model}
\end{equation}
where the transition rates are written above the arrows. $g_{\pm}$ are dimensionless functions, chosen so that for any $y$: $g_+(y)+g_-(y)=1$; $\lambda$ is a rate. 
Notice $\bar{\rho}_{N,R}(x,t)$ is dimensionless and typically of order $1$, and can be written as a convolution: $\bar{\rho}_{N,R}(x,t)=\int \chi_R(x-y) \hat{\rho}_N(y)dy$, with 
\[
\chi_R(x)=\frac{1}{2R}\chi\left(\frac{x}{R}\right)~,~ \hat{\rho}_N(x)=\frac{L}{N} \sum_{n=1}^N \delta(x-x_n).
\]
Writing $x_i(t),v_i(t)$ for the position and velocity of particle $i$  ($v_i=\pm v$), the positions dynamics is simply $\dot{x}_i = v_i$.  
One possible example for the functions $g_\pm$ is: $g_+(\rho)=1/(1+b\rho)$; $g_-(\rho)=b\rho/(1+b\rho)$, for some parameter $b>0$.

This model has three length scales: the interaction range $R$, the system's size $L$ and the mean free path
\[
\ell = \frac{v}{\lambda}.
\]
We assume in the following $\ell  \ll  L$, and we will make hypotheses on $R$ later on. 

\paragraph{Kinetic equation.}

In an appropriate large $N$ limit, this microscopic model can be described by a kinetic equation. 
We use from now on the typical time between velocity switches $\lambda^{-1}$ as time unit and the mean free path $\ell$ as space unit, so that the particles' velocities are $\pm 1$. 
We define the empirical phase space densities 
\begin{equation}
\hat{f}_{N,\pm} = \frac{L}{N \ell} \sum_{i=1}^N \delta\big(x-x_i(t)\big)\delta_{v_i(t),\pm 1}
\end{equation}
where $L$ is the size of the system. $\varepsilon^{-1}=N\ell/L$ is the typical number of particles in an interval of size the mean free path; we assume this number to be large. With this normalization, the phase space densities $f_{N,\pm}$ typically take values of order $1$.  

Using test functions varying over lengthscales at least $\ell$, or equivalently after coarse-graining on a scale $\ell$, the empirical densities $\hat{f}_{N,\pm}$ approach in the limit $\varepsilon \to 0$ the limiting functions $f_\pm$, which satisfy the kinetic equation: 
\[
\begin{cases}
\partial_{t}f_{+}+ \partial_{x}f_{+} & =-g_-\left(\chi_R \ast \rho\right)f_{+}+g_{+}\left(\chi_R \ast \rho\right) f_-\\
\partial_{t}f_{-} -\partial_{x}f_{-} & =-g_+\left(\chi_R \ast \rho\right)f_{-}+g_{-}\left(\chi_R \ast \rho\right)f_+
\end{cases},
\]
where $\rho(x,t)=f_+(x,t)+f_-(x,t)$; $R$ is now expressed in unit of the mean free path $\ell$. Introducing $M^R_\pm[\rho](x)= g_\pm( (\chi_R \ast \rho)(x)) \rho(x)$, the kinetic equation reads
\[
\begin{cases}
\partial_{t}f_{+}+ \partial_{x}f_{+} & =-f_+(x)+M^R_+[\rho](x) \\
\partial_{t}f_{-} -\partial_{x}f_{-} & =-f_-(x)+M^R_-[\rho](x).
\end{cases}
\]
Densities $f_\pm$ such that $f_+=M_+^R[\rho]$ and $f_-=M_-^R[\rho]$ are local equilibria. Instead of $f\pm$, it will often be convenient to use the fields $\rho$ and $m=f_+-f_-$: $m$ describes the local average polarization of the particles.

\paragraph{Large deviations.}

Since the microscopic dynamics is a jump process, we can also characterize dynamical fluctuations
of the empirical phase space measure with the following large deviation principle, as done for instance in \cite{Rezakhanlou,FreddyBoltzmann,OuassimSOH}. Details are given in Appendix \ref{sec:appendixA}. 
\begin{equation}
\mathbb{P}\left(\left\{ f_{N,+}(t),f_{N,-}(t)\right\} _{0\leq t\leq T}\simeq \left\{ f_{+}(t),f_{-}(t)\right\} _{0\leq t\leq T}\right)\underset{N\rightarrow+\infty}{\asymp}\text{e}^{-\varepsilon^{-1}I_{T}^{\rm kin}[f]},\label{eq:LDPkinetic}
\end{equation}
\begin{equation}
I_{T}^{\rm kin}[f]=\int_{0}^{T}\text{d}t\, \sup_{p} \left\{\int_0^L \big(p_{+}\partial_{t}f_{+}+p_{-}\partial_{t}f_{-}\big)\text{d}x-H\left[f,p\right]\right\} \label{eq:Kinetic_ratefunction}
\end{equation}
where  
\[
H\left[f,p\right]=H_{T}\left[f,p\right]+H_{J}\left[f,p\right].
\]
Here $f=\left(f_{+},f_{-}\right)$, $p=\left(p_{+},p_{-}\right)$ are the momentum fields conjugate to $f_+,f_-$, $H_T$ is the "transport" Hamiltonian
\begin{equation}
H_{T}\left[f,p\right]=\int_0^L\left(-p_{+}\partial_{x}f_{+}+p_{-}\partial_{x}f_{-}\right) \text{d}x,
\label{eq:HT}
\end{equation}
and $H_J$ is the "jump" Hamiltonian
\begin{equation}
H_{J}\left[f,p\right]=\int_0^L \left( \frac{f_{+}M_{-}^R}{ \rho}\left(\text{e}^{-p_{+}+p_{-}}-1\right)+\frac{f_{-}M_{+}^R}{ \rho}\left(\text{e}^{-p_{-}+p_{+}}-1\right)\right)\text{d}x.
\label{eq:HJ}
\end{equation}
If the initial condition is random, there are also fluctuations associated to it, and a corresponding rate function. For particles at $t=0$ which are independent, the large deviation speed is also $\varepsilon^{-1}$.  In Eq.\eqref{eq:LDPkinetic}, $\left\{ f_{N,+}(t),f_{N,-}(t)\right\} \simeq \left\{ f_{+}(t),f_{-}(t)\right\}$ has the informal meaning "$\left\{ f_{N,+}(\cdot),f_{N,-}(\cdot)\right\}$ is in a small neighborhood of $\left\{ f_{+}(\cdot),f_{-}(\cdot)\right\}$".
The large deviation principle can be expressed in $\rho,m$ variables. The Hamiltonian reads (keeping for simplicity the notations $H_T$ and $H_J$): 
\begin{eqnarray}
H_{T}\left[\rho,m,p_\rho,p_m\right]&=&-\int_0^L\left(p_m\partial_{x}\rho+p_\rho\partial_{x}m\right) \text{d}x,\\
H_{J}\left[\rho,m,p_\rho,p_m\right]&=& \int_0^L \left( \frac{(\rho+m)M_{-}^R}{ 2\rho}\left(\text{e}^{-2 p_{m}}-1\right)+\frac{(\rho-m)M_{+}^R}{ 2\rho}\left(\text{e}^{2p_{m}}-1\right)\right)\text{d}x,
\label{eq:HJ_full}
\end{eqnarray}
where we introduced the associated conjugate momentum fields  $p_\rho = (p_+ + p_-)/2$ and $p_m = (p_+ - p_-)/2$.
$H_J$ clearly does not depend on $p_\rho$, which indicates that the equation for $\rho$ is deterministic and does not give rise to fluctuations.

\paragraph{Macroscopic scaling.}
We are interested in situations where the size of the system $L$ is much larger than the mean free path $\ell$. Hence it makes sense to 
consider profiles $f_\pm$ which vary over length and time scales much larger than $\ell$ and $\lambda^{-1}$. We then introduce the Knudsen number $\alpha=\ell/L$, where $L$ is the macroscopic length, 
and rescale space and time as $\tilde{t}=\alpha t,\,\tilde{x}=\alpha x$: this is a hyperbolic rescaling.
The kinetic equation becomes (removing the tildes for an easier reading):
\begin{equation}
\alpha \left(\partial_{t}f_{\pm}\pm\partial_{x}f_{\pm} \right) =-f_{\pm}+M_{\pm}^R [\rho].
\label{eq:kinetic2-1}
\end{equation}
We also assume now that the interaction range $R$ is small, at most of order of the mean free path. Since $R$ is now expressed in macroscopic units, it is at most of order $\alpha$. Hence a simple Taylor expansion, and the identity $\int x\chi(x) dx=0$ yields 
\[
(\chi_R \ast \rho)(x) = \rho(x) + O(\alpha^2).
\]
Neglecting the $O(\alpha^2)$ terms in the small $\alpha$ limit, we may then replace $\chi_R \ast \rho$ by $\rho$, and the kinetic equation becomes
\begin{equation}
\alpha \left(\partial_{t}f_{\pm}\pm\partial_{x}f_{\pm} \right) =-f_{\pm}+M_{\pm}(\rho),
\label{eq:kinetic2}
\end{equation}
where $M_\pm=g_\pm(\rho(x)) \rho(x)$: $M_\pm$ are now local functions of $\rho$.
Using the $\rho,m$ fields, and $a(\rho)= M_+(\rho)-M_-(\rho)$, \eqref{eq:kinetic2} reads
\begin{eqnarray}
\partial_t \rho +\partial_x m &=& 0 \nonumber \\
\alpha\left(\partial_t m +\partial_x \rho\right) &=& a(\rho)-m.
\label{eq:kinetic3}
\end{eqnarray}
We shall now use the $\rho,m$ variables.
The kinetic equation \eqref{eq:kinetic2} or \eqref{eq:kinetic3} is the starting point for many mathematical works (see for instance \cite{Natalini98,Aregba}), which are often interested in proving the hydrodynamic (i.e. $\alpha \to 0$) limit in an appropriate scaling, and sometimes look for efficient numerical methods to simulate conservation laws. 
From \eqref{eq:kinetic3} one then expects the field $m$ to be close to $a(\rho)$ in the small $\alpha$ limit.
At leading order in $\alpha$, one then obtains the equation for $\rho$:
\begin{equation}
\partial_t \rho +\partial_x\big(a(\rho)\big) = 0.
\label{eq:conservation}
\end{equation}
This is a nonlinear conservation law. Since it plays the role of Euler equation in standard hydrodynamics, we shall sometimes call it in the following "Euler equation". An expansion in the parameter $\alpha$ would add a diffusive term of order $\alpha$ to \eqref{eq:conservation}, and turn it into 
a one-dimensional analogue of compressible Navier-Stokes equations. We shall follow this route in section \ref{sec:-convergence-neglecting-shocks}, at the level of large deviations.

Under this hyperbolic rescaling and using the $\rho,m$ variables, the rate function (\ref{eq:Kinetic_ratefunction}) reads (with $\tilde{T}=\alpha T$):
\begin{equation}
I_{T}^{{\rm kin},\alpha}[\rho,m]=\alpha^{-2}\int_{0}^{\tilde{T}}\text{d}\tilde{t}\,\sup_{p}\bigg\{ \int_0^1 \text{d}\tilde{x}\, \alpha\big(p_{\rho}\partial_{\tilde{t}}\rho +p_{m}\partial_{\tilde{t}}m +p_{m}\partial_{\tilde{x}}\rho+p_{\rho}\partial_{\tilde{x}}m \big)
- H_J[\rho,m,p_\rho,p_m]\bigg\}. \label{eq:rate_function2a}
\end{equation}
Neglecting terms of order $\alpha^2$,  we replace $M_\pm^R$, which contain convolutions, by the local quantities $M_\pm$ in the expression of $H_J$. If $R\ll \ell$, the terms neglected are even smaller; we shall comment on these at the end of section \ref{sec:gen_comp}. 

The Hamiltonian $H_J$ contains a part linear in $p$ which contributes to the deterministic kinetic equation \eqref{eq:kinetic3}, and a part of order quadratic in $p$ or higher which describes the stochastic 
fluctuations around \eqref{eq:kinetic3}. It is convenient to separate both terms, so we rewrite the rate function (removing the $\tilde{}$ for an easier reading):
\begin{multline}
I_{T}^{{\rm kin},\alpha}[\rho,m]=\alpha^{-2}\int_{0}^{T}\text{d}t \sup_{p}\bigg\{\int_0^1\text{d}x\,\big[\alpha\left(p_{\rho}\partial_{t}\rho+p_{m}\partial_{t}m+p_{m}\partial_{x}\rho+p_{\rho}\partial_{x}m\right)  \\
+(m-a(\rho))p_{m}\big]  - H_Q[\rho,m,p_\rho,p_m]\bigg\}, \label{eq:rate_function2}
\end{multline}
where we have defined $H_Q$ by removing the linear in $p$ part of $H_J$ ($Q$ stands for "quadratic"):
\[
 H_Q[\rho,m,p_\rho,p_m]  = H_J[\rho,m,,p_\rho,p_m]-\int \big[ (m-a(\rho))p_{m}\big].
\]
Under the hyperbolic rescaling, the large deviation speed for the initial fluctuation becomes $(\varepsilon \alpha)^{-1}$, whereas \eqref{eq:rate_function2} seems to correspond to a large deviation speed $\varepsilon \alpha^{-2}$. 

\paragraph{Quadratic approximation.}
\label{sec:quadratic}
The microscopic jump process naturally creates a Poissonian noise, expressed by the Hamiltonian \eqref{eq:HT}. In the macroscopic scaling however, this Poissonian noise is often replaced by a 
Gaussian approximation, which amounts to use a quadratic approximation to the Hamiltonian. See \cite{Agranov21,Agranov22} for a discussion in the context of active lattice gases and \cite{OuassimSOH} in the context 
of an off lattice active matter model.  

This quadratic approximation for the Hamiltonian is valid as long as the momentum $p_m$ remains small ($H_J$ does not depend on $p_\rho$). For a given profile $(\rho,m)$, $p_m$ solves the equation
\[
\alpha(\partial_t m +\partial_x \rho)+m-a(\rho) = \frac{\delta H_Q}{\delta p_m},
\]

where the leading order at small $p_m$ on the right hand side is linear in $p_m$. From this equation we see that in the small $\alpha$ regime, $p_m$ remains small under the two following hypotheses:
\begin{enumerate}
\item $m$ is close to $a(\rho)$: this amounts to consider profiles which are close to local equilibrium;
\item $\alpha (\partial_t m + \partial_x \rho)$ is small: this amounts to consider profiles which do not vary too steeply in space and time.
\end{enumerate}
In the following, we shall use this quadratic approximation for the Hamiltonian and discuss its validity; in particular, we can check with the above criteria that our computations are consistent.
The jump Hamiltonian truncated at quadratic order reads:
\begin{equation}
H_{J}\left[\rho,m,p_\rho,p_m\right]=\int \bigg[ -(m-a(\rho)) p_{m} +\frac{1}{\rho}\left(\rho^2-ma(\rho)\right)p_m^2 \bigg] dx.
\label{eq:quad2}
\end{equation}

\paragraph{Lagrangian expression.}
Under the quadratic approximation for the Hamiltonian \eqref{eq:quad2}, it is easy to solve the optimization on $(p_\rho,p_m)$, and to obtain a Lagrangian expression for the kinetic rate function.
Maximizing over $p_\rho$ ensures that the deterministic equation for $\rho$ appears as a constraint. Maximizing over $p_m$ then yields the rate function:
\begin{eqnarray}
I_{T}^{{\rm kin},\alpha}[\rho,m]&=&\frac{\alpha^{-2}}{4}\! \int_{0}^{T}\!\!\!\text{d}t\int_0^1\!\!\text{d}x\,\frac{\rho}{\rho^2 -m a(\rho)}\left\{ \alpha\left(\partial_{t}m+\partial_{x}\rho\right)+m-a\right\} ^{2}~{\rm if}~\partial_{t}\rho+\partial_{x}m=0, \label{eq:lagrangian}\\ 
&=& +\infty ~{\rm otherwise}.
\label{eq:lagrangian2}
\end{eqnarray}
This kinetic LDP (\ref{eq:lagrangian}-\ref{eq:lagrangian2}) encodes the finite size fluctuation of the system at kinetic level around the kinetic equation \eqref{eq:kinetic2}. It is our starting point in the following sections.

\section{Fluctuating hydrodynamics from the kinetic LDP (without shocks)}

\label{sec:-convergence-neglecting-shocks} 

In this section, we start from the kinetic large deviation rate function \eqref{eq:lagrangian}, obtained in the hyperbolic scaling and in the quadratic approximation. Our goal is to derive
a large deviation principle for the density field $\rho$
in the small $\alpha$ limit. At the deterministic level, we have seen that one obtains in the small $\alpha$ limit a Euler-like equation at leading order \eqref{eq:conservation}. The next to leading order in $\alpha$ is a Navier-Stokes-like equation, which includes a diffusion term of order $O(\alpha)$.
From \eqref{eq:lagrangian}, we want to compute both this diffusion term and the stochastic fluctuations around this Navier-Stokes-like equation.
To be more precise, we fix a density profile $\rho(x,t)$, which we assume to be regular. Our goal is to compute at the large deviation level and in the small $\alpha$ limit the probability to observe $\rho$. In other words, we want to understand the small $\alpha$ behavior of
\[
\mathcal{I}^\alpha_T[\rho] = \lim\limits_{\varepsilon \to 0} \left[-\varepsilon \mathbb{\log P}\big(\rho,m \big)\right].
\] 
It is given by a contraction principle:
\begin{equation}
\lim\limits_{\varepsilon \to 0}\left[ -\varepsilon\mathbb{\log P}\left( \rho,m \right)\right] = \underset{m}{\inf}I_{T}^{{\rm kin},\alpha}[\rho,m].
\label{eq:contraction}
\end{equation}

Let us first assume that
\begin{equation}
\partial_t \rho +\partial_x\big(a(\rho)\big) = O(1).
\label{eq:not_approx_sol}
\end{equation}
In other words, $\rho$ is far from being an approximate solution of Euler equation \eqref{eq:conservation}. Then it is not possible to choose $m$ to be close to the local equilibrium $m=a(\rho)$: indeed the constraint $\partial_{t}\rho+\partial_{x}m=0$ would impose $ \partial_t \rho +\partial_x\big(a(\rho)\big)$ to be small. We conclude that the optimal $I_{T}^{{\rm kin},\alpha}[\rho,m]$ is in this case of order $\alpha^{-2}$, corresponding to a large deviation speed $\varepsilon^{-1} \alpha^{-2}$. This implies that at the slower large deviation speed $(\varepsilon \alpha)^{-1}$ (the speed of initial condition large deviations) $\rho$ must be approximately a solution of Euler equation \eqref{eq:conservation}. We have recovered the well known fact that, at these space and time scales, fluctuations are dominated by the initial conditions. This result is at the basis of "Ballistic Macroscopic Fluctuation Theory" \cite{Doyon19,Doyon22a,Doyon22b}, and crucially relies on the absence of shocks, as will become clear in Section \ref{sec:JV}.

We now assume that $\rho$ is approximately a solution of Euler equation \eqref{eq:conservation}:
\begin{equation}
\partial_t \rho +\partial_x\big(a(\rho)\big) = O(\alpha).
\label{eq:approx_sol}
\end{equation}
It is then possible to choose $m$ to be close to the local equilibrium $m=a(\rho)$, and we can anticipate that the optimal $I^{{\rm kin},\alpha}_{T}[\rho,m]$ is much smaller than $\alpha^{-2}$.  We conclude that density profiles $\rho$ that are approximate solutions of the Euler equation \eqref{eq:conservation} are in the small $\alpha$ limit overwhelmingly more probable than density profiles which are not. We now choose such a profile $\rho$ and compute the leading order of $\mathcal{I}^\alpha_{T}[\rho]$.

Define $g$ such that 
\begin{equation}
m=a(\rho)+\alpha g +\mathcal{O}\left(\alpha^{2}\right),\label{eq:hydro_profile}
\end{equation}
At leading order in $\alpha$, we can rewrite the rate function 
\[
I_{T}^{{\rm kin},\alpha}[\rho,m]=\int_{0}^{T}\text{d}t\int_0^1\text{d}x\,\frac{\rho}{4(\rho^2-a(\rho)^2)}\big\{ \big(1-a'(\rho)^2\big)\partial_{x}\rho+ g \big\} ^{2},
\]
with the constraint 
\[
\partial_{t}\rho+\partial_{x}\left(a\left(\rho\right)\right)+\alpha\partial_{x}g=0,
\]
Now, instead of minimizing over $m$, we minimize over $g$.
We further split $g$ in a deterministic and a stochastic part: $g=g^d+g^s$, where the deterministic part is fixed
\[
g^{d}=- D(\rho)\partial_x \rho,
\]
with the diffusion coefficient
\begin{equation}
D(\rho)=\big(1-a'(\rho)^2\big).
\label{eq:D}
\end{equation}
We still have to optimize over the stochastic part $g^{s}$.
The constraint can finally be written:
\[
-\alpha g^s = \int^x \big[\partial_t \rho+\partial_x a(\rho)-\alpha\partial_x \big(D\partial_x \rho\big)\big]+{\rm cst}
\]
where $\int^x u$ denotes a primitive of $u$.
All in all, we can compute the leading order of the rate function as $\alpha$ goes to zero and
the final result reads 
\begin{equation}
\mathcal{I}^\alpha_{T}\left[\rho\right]= \alpha^{-2}\int_{0}^{T}\text{d}t\int_0^1\text{d}x\,
\frac{\rho}{4\big(\rho^2-a(\rho)^2\big)}\left\{ \int^{x}\left(\partial_{t}\rho+\partial_{x}\left(a\left(\rho\right)\right)-\alpha\partial_{x}\left(D\left(\rho\right)\partial_{x}\rho\right)\right)\right\} ^{2}, \label{eq:LDPalpha}
\end{equation}
where the constant in the primitive is chosen so as to minimize $\mathcal{I}^\alpha_{T}$.

\eqref{eq:LDPalpha} recovers \eqref{eq:ldf_diff} with the function
$D$ given by \eqref{eq:D}, and
\[
\sigma(\rho)= \frac{2(\rho^2-a^2)}{\rho}.
\]
 All the relevant microscopic information is then contained in a few macroscopic quantities $a$, $D$ and $\sigma$, which is a crucial ingredient in macroscopic fluctuation theory.  
We see from \eqref{eq:LDPalpha} that if $\rho$ is regular and satisfies \eqref{eq:approx_sol} (i.e. is an approximate solution of the conservation law), $\mathcal{I}^\alpha_{T}[\rho]$ is of order $1$. \eqref{eq:LDPalpha} makes sense also if $\rho$ satisfies \eqref{eq:not_approx_sol}, i.e. is \emph{not} an approximate solution of the conservation law. We also recover in this case from \eqref{eq:LDPalpha} that 
 $\mathcal{I}^\alpha_{T}$ is of order $\alpha^{-2}$. However, we must keep in mind that although this order of magnitude is correct, expression \eqref{eq:LDPalpha} does {\it not} describe the large deviations when $\rho$ is not an approximate solution of the conservation law: indeed the criteria of section\ref{sec:quadratic} for using the quadratic Hamiltonian are not satisfied in this case.

A stochastic PDE for $\rho $ with the required large deviation behavior \eqref{eq:LDPalpha} would read 
\begin{equation}
\partial_{t}\rho +\partial_{x}\left(a\left(\rho \right)\right)=\alpha\partial_{x}\left(D\left(\rho \right)\partial_{x}\rho \right)+
\partial_{x}\left(\sqrt{\varepsilon \alpha^2 \sigma(\rho)}\,\eta\right).
\label{eq:spde}
\end{equation}
\eqref{eq:LDPalpha} and \eqref{eq:spde} are typical large deviation rate function and stochastic PDE for diffusive systems with simultaneously vanishing diffusion and noise; this is the context of the work \cite{Mariani10}.
The diffusion coefficient \eqref{eq:D} can in principle become negative. In this case the solution of the kinetic system \eqref{eq:kinetic3} does not even converge to the solution of the conservation law \eqref{eq:conservation} in the $\alpha\to 0$ limit, see \cite{Natalini98} condition (1.15). $D(\rho)>0$ has a clear physical interpretation: the characteristic speed $|a'(\rho)|$ should not be larger than the microscopic speed $1$ (see \cite{Natalini98} (2.14)).

We have used in this computation the quadratic approximation to the Hamiltonian. We may ask if the conditions to use this approximation are fulfilled. Clearly, as explained above, they are not when $\rho$ satisfies \eqref{eq:not_approx_sol}. 
However, when $\rho$ satisfies \eqref{eq:approx_sol}, 
we have seen that the optimal "magnetization" $m$ is close to local equilibrium, so the first condition can be met. The second condition requires that the space and time derivatives of $\rho$ remain of order $1$, 
or at least smaller than $\alpha^{-1}$. This is actually a very stringent condition put on the profiles $\rho$, because solutions of the conservation law \eqref{eq:conservation} typically develop shocks.
Hence we recover for the validity of the large deviation rate function \eqref{eq:LDPalpha} the same type of limitation that have long been identified for the rigorous derivation of hydrodynamic equations of the compressible Navier-Stokes type, see for instance \cite{Caflisch80,BGL91}

\section{Jensen--Varadhan large deviation functional}

\label{sec:JV} 
Section \ref{sec:-convergence-neglecting-shocks} provides the large deviation speed and rate function for density profiles which are approximate solutions of Euler equation (the conservation law) and are regular enough. 
We now assume the results of Section \ref{sec:-convergence-neglecting-shocks} are valid also for steep density profiles, and use them to compute the weight of shocks and antishocks when $\alpha\to 0$.
We know it may not be correct, but it will provide a comparison with the direct computations which follow in section~\ref{sec:direct}.

We then fix an antishock $\rho^A$, and look for the best way to approximate $\rho^A$ by a family of profiles $\rho^\alpha$, where "best" means "the most probable", or rather "least improbable" way. We want to compute 
\begin{equation}
\inf_{\rho^\alpha \to \rho^A} \lim\limits_{\alpha \to 0}\alpha^\gamma \mathcal{I}_{T}^\alpha\left[\rho^\alpha \right], \label{eq:limit}
\end{equation}
where $\mathcal{I}_{T}^\alpha$ is given by \eqref{eq:LDPalpha}, and $\gamma$ is chosen so that the $\alpha \to 0$ limit is non trivial.
Technically, this is a $\Gamma$-convergence result for the rate
function \eqref{eq:LDPalpha}. This computation has been done by Mariani et al. \cite{Mariani10,BBMN}, who provide an almost complete proof of the $\Gamma$-convergence towards a generalized 
Jensen-Varadhan's functional. We shall recall and make use of these results later on.


Let $\rho^{A}$ be an antishock between $\rho_{L}$ and $\rho_{R}$,
with speed given by the Rankine-Hugoniot relation: $v_{s}=\left(a\left(\rho_{R}\right)-a\left(\rho_{L}\right)\right)/\left(\rho_{R}-\rho_{L}\right)$. $a$ convex implies $\rho_L<\rho_R$ for an antishock. We look for
a family $(\rho^{\alpha})$ of regularized profiles approximating
$\rho^{A}$ in the $\alpha\to0$ limit in an optimal way. This is the situation represented on Fig.~\ref{fig:shock}. In principle, we should complement this antishock by a shock between $\rho_{R}$ and $\rho_{L}$, traveling at the same speed, in order to accommodate the periodic boundary condition. However, since all shock related length scales are much smaller than the macroscopic one, the shock can be dealt with independently. And as it will become clear, it does not contribute to the rate function.

First assume that $(\rho^{\alpha})$ regularizes $\rho^A$ over a spatial scale $\ell_{\rm reg}$. Then far from the discontinuity of $\rho^A$, $\rho^\alpha$ is uniform and does not contribute to the rate function \eqref{eq:LDPalpha}.
In a window of width $\ell_{\rm reg}$ around the discontinuity, $\partial_t \rho^\alpha+\partial_x a(\rho^\alpha)$ and $\alpha \partial_{x} (D \partial_{x}\rho^\alpha)$ are of order $\ell_{\rm reg}^{-1}$ and $\alpha \ell_{\rm reg}^{-2}$ respectively. 
Thus, if $\ell_{\rm reg} \gg \alpha$, the term $\partial_t \rho^\alpha+\partial_x a(\rho^\alpha)$ dominates. 
Taking the primitive as in \eqref{eq:LDPalpha} brings a factor $\ell_{\rm reg}$; integrating over space brings another factor $\ell_{\rm reg}$, since the integrand is non negligible only within a window of size $\ell_{\rm reg}$ around the antishock. Hence \eqref{eq:LDPalpha} is of order $\alpha^{-2}\ell_{\rm reg}\gg \alpha^{-1}$. 
If on the other hand $\ell_{\rm reg} \ll \alpha$, the term $\alpha \partial_{x} (D \partial_{x}\rho^\alpha)$
dominates and \eqref{eq:LDPalpha} is of order $\ell_{\rm reg}^{-1}\gg \alpha^{-1}$.
We conclude that the order of magnitude of \eqref{eq:LDPalpha} is smallest when $\ell_{\rm reg} \propto \alpha$, and in this case it is $\alpha^{-1}$. In other words, we expect that the optimal
spatial scale over which $\rho^{\alpha}$ regularizes the antishock $\rho^A$ is $\alpha$, in the sense that any smoother or steeper regularization corresponds to a much larger value of the rate function. We have assumed here that the diffusion $D$ and macroscopic velocity $a'$ are of order $1$. Since we expect that the optimal value of $\mathcal{I}_{T}^\alpha$ is of order $\alpha^{-1}$, we shall then take $\gamma=1$ so that the limit in \eqref{eq:limit} is well defined. 

Since $\rho^{\alpha}$ is a traveling antishock regularized over a length scale of order $\alpha$, we may define a negative diffusion coefficient $-\tilde{D}[\rho^\alpha]$ through the following equation:
\begin{equation}
\partial_{t}\rho^{\alpha}+\partial_{x}\left(a\left(\rho^{\alpha}\right)\right)=-\alpha\partial_{x}\left(\tilde{D}[\rho^\alpha]\partial_{x}\rho^{\alpha}\right)~,~\lim_{x\to -\infty} \rho^\alpha(x,t) =\rho_L,~\lim_{x\to +\infty} \rho^\alpha(x,t). =\rho_R\label{eq:Dtilde1}
\end{equation}
Recall that a positive diffusion coefficient allows to approximate only shock profiles in the small diffusion limit. Alternatively, we may define $\rho^\alpha$ as the space-time reversal of the solution to Eq.~\eqref{eq:Dtilde1} with a positive diffusion (and switched boundary conditions at $\pm \infty$). The notation $\tilde{D}[\cdot]$ means that $\tilde{D}$ depends functionally on $\rho^\alpha$.
From now on we consider that the sequence of profiles $\rho^\alpha$ is parameterized by the sequence of diffusion coefficients $\tilde{D}$.
The goal is now to find the best possible sequence $\rho^{\alpha}$, or equivalently the best possible $\tilde{D}$. 

With these new notations, the rate function \eqref{eq:LDPalpha} reads 
\begin{equation}
\mathcal{I}_{T}^{\alpha}\left[\rho^{\alpha}\right]= \frac{1}{2} \int_0^T \text{d}t \int \text{d}x\,\frac{1}{\sigma\left(\rho^{\alpha}\right)}\left(\left(\tilde{D}\left[\rho^{\alpha}\right]+D\left(\rho^{\alpha}\right)\right)\partial_{x}\rho^{\alpha}\right)^{2}.
\label{eq:rate_inter}
\end{equation}
We have to optimize it over $\tilde{D}[\cdot]$, which is a priori a difficult task. Following similar computations in the context of exclusion processes \cite{BodineauDerrida2006}, the problem \eqref{eq:limit} is solved in \cite{Mariani10,BBMN} (a heuristic version of this computation is provided for completeness in Appendix \ref{sec:appendixB}). Their results can be summarized as follows:
\begin{itemize}
\item The relevant scaling is $\gamma=1$, as found heuristically above.
\item The infimum over $\tilde{D}[\cdot]$ in the variational problem \eqref{eq:rate_inter} is reached precisely when the functional $\tilde{D}[\rho]$ is the local function $D(\rho)$. This will be useful in the following section.
\item Inserting the traveling wave profile approximating the antishock into \eqref{eq:rate_inter}, one obtains the sought infimum, which is the following generalized Jensen-Varadhan functional:
\begin{equation}
I_{JV}\left[\rho^{A}\right]= \inf_{\rho^\alpha \to \rho^A} \lim\limits_{\alpha \to 0}\alpha \mathcal{I}_{T}^\alpha\left[\rho^\alpha \right]= T \int_{{\rm min}(\rho_{L},\rho_R)}^{{\rm max}(\rho_L,\rho_{R})}\text{d}u\,\frac{2 D(u)}{\sigma \left(u\right)}\left[\big(a\left(\rho_{L}\right)-a(u)-v_{s}(\rho_L-u)\big){\rm sgn}(\rho_R-\rho_L)\right]_+,\label{eq:JVLDF}
\end{equation}
where $[C]_+$ is the positive part of $C$, and the expression \eqref{eq:JVLDF} is also valid for a shock (entropic solution). 
A few comments are in order. For $a$ concave or convex and $\rho^A$ an antishock, the quantity $[-v_{s}\rho_{L}+a\left(\rho_{L}\right)+v_{s}u-a(u)]{\rm sgn}(\rho_R-\rho_L)$ is positive for any $u$ between $\rho_L$ and $\rho_R$. If the flux $a$ has not a definite concavity, this quantity can change sign. A shock (entropic solution) corresponds to the case where it is negative for all $u$ between $\rho_L$ and $\rho_R$; then, 
thanks to the positive part, the Jensen-Varadhan functional \eqref{eq:JVLDF} vanishes, as it should for a deterministic solution.
\end{itemize}

The infimum in \eqref{eq:rate_inter} is then reached for a sequence $(\rho^\alpha)$ which is the space-time reversal of the (deterministic) evolution equation \eqref{eq:spde} without noise. This fact may seem surprising since the system is microscopically not reversible. Some reversibility then seems to be restored at the macroscopic scale, which may also be seen in the fact that (unless there is some boundary forcing) the quasi-potential $\int s(\rho) dx$ is local, and satisfies an Einstein relation $s^{''}(\rho)=D(\rho)/\sigma(\rho)$  \cite{Bellettini10}; it has indeed been shown that in more than one dimension, when the matrices $D$ and $\sigma$ are not proportional, the structure of the optimal antishock may be more complicated (see \cite{BBC}, conclusion section). 

We end this section by two further comments. First, although \eqref{eq:JVLDF} is written for a single antishock, it extends easily to profiles containing several antishocks, or whose discontinuity varies with time: the contributions of each antishock simply add up.
\eqref{eq:JVLDF} is then the basic building block to compute the cost of any weak solution. Furthermore, starting from a smooth profile, weak solutions can indeed create antishocks; this is used in \cite{Bellettini10,Bahadoran10} to compute quasi-potentials.  

Second, the $\gamma=1$ scaling implies that the large deviation speed for an antishock is $(\varepsilon \alpha)^{-1}$. Hence antishocks appear at the same level as fluctuations of initial conditions. Noticeably, this forces to modify Ballistic Macroscopic Fluctuation Theory in this context \cite{Doyon19,Doyon22a}. 

\section{Weight of an antishock directly from the kinetic LDP}

\label{sec:direct}
 In section \ref{sec:-convergence-neglecting-shocks} we have obtained, under appropriate hypotheses, the rate function for the density at the diffusive (Navier-Stokes) level, for small but finite Knudsen number $\alpha$; then in section \ref{sec:JV} we have taken the limit  $\alpha\to 0$. The hypotheses in section \ref{sec:-convergence-neglecting-shocks}
require in particular that the optimal kinetic function $f$ remains close to a local equilibrium of the kinetic equation, which is a priori not guaranteed for an antishock when $\alpha\to 0$. 
In order to shed light on the validity of the rate function \eqref{eq:JVLDF}, we should start directly from the non quadratic kinetic rate function \eqref{eq:Kinetic_ratefunction} and take the $\alpha\to 0$ limit. This is very difficult however. Hence in this section
we shall take the $\alpha \to 0$ limit starting from the quadratic kinetic rate function \eqref{eq:lagrangian}, keeping in mind the hypotheses for its validity are not necessarily verified: indeed they will not be in Sec. \ref{sec:gen_comp}.
We want to compute the probability of an antishock. Technically, as in section \ref{sec:-convergence-neglecting-shocks}, we need
to perform a contraction from the kinetic functions "$(\rho,m)$-space" to the density functions "$\rho$-space", with the extra difficulty
that the target density $\rho$ is not regular. We show that even starting from the quadratic kinetic rate function \eqref{eq:lagrangian}, there is no reason that the optimal kinetic evolution $\rho,m$ corresponding
to the antishock remains close to a local equilibrium, and no reason
that the Jensen-Varadhan's functional \eqref{eq:JVLDF} remain valid.

In general, the large deviation rate function replacing \eqref{eq:JVLDF} that
we obtain is not very explicit. However, we also identify the regime where \eqref{eq:JVLDF} is valid: it corresponds to the cases where the length over which shocks and antishocks are regularized is large with respect to the mean free path; it also corresponds to a small Mach number regime.

\subsection{The general computation}
\label{sec:gen_comp}
The starting point is the kinetic rate function \eqref{eq:lagrangian}-\eqref{eq:lagrangian2} which we recall here:
\begin{eqnarray}
I_{T}^{{\rm kin},\alpha}[\rho,m]&=&\frac{\alpha^{-2}}{4}\! \int_{0}^{T}\!\!\!\text{d}t\int_0^1\!\!\text{d}x\,\frac{\rho}{\rho^2 -m a(\rho)}\left\{ \alpha\left(\partial_{t}m+\partial_{x}\rho\right)+m-a\right\} ^{2}~{\rm if}~\partial_{t}\rho+\partial_{x}m=0, \label{eq:Kldp}\\ 
&=& +\infty ~{\rm otherwise}.
\label{eq:constraint}
\end{eqnarray}
Let $\rho^{A}$ be an antishock between $\rho_{L}$ and $\rho_{R}$ (we assume for simplicity that $\rho_L<\rho_R$, which is the case when $a$ is convex on $[\rho_L,\rho_R]$),
with speed $v_{s}=\left(a\left(\rho_{R}\right)-a\left(\rho_{L}\right)\right)/\left(\rho_{R}-\rho_{L}\right)$. As in the previous section, we disregard the boundary conditions here, which should be accommodated by a shock.
In principle, our goal is to perform a contraction from the kinetic
LDP, ie to compute (the $\alpha$ factor enforces the same scaling as in Sec.\ref{sec:JV})
\[
\mathcal{J}_T(\rho^A) = \inf_{m}\left(\alpha I_{T}^{{\rm kin},\alpha}[\rho^A,m]\right).
\]
However, $\rho^{A}$ is singular; thus the possible $m$
would be singular too, and would correspond to an infinite value for
the rate function. We then introduce $(\rho^{\alpha})$ a family of
regularized profiles approximating $\rho^{A}$ in the $\alpha\to0$
limit, to be determined. As in section \ref{sec:JV}, we parameterize the family $(\rho^{\alpha})$ by the diffusion coefficient
$\tilde{D}[\rho]$ (functional of $\rho$), through the equation 
\begin{equation}
\partial_t \rho^\alpha +\partial_x (a(\rho^\alpha)) = -\alpha \partial_x \left( \tilde{D}[\rho^\alpha]\partial_x \rho^\alpha \right). \label{eq:Dtilde}
\end{equation}
Since $\rho^\alpha$ regularizes an antishock, the diffusion is negative, as seen in the previous section. Furthermore, $\rho^\alpha$ regularizes the antishock $\rho^A$ over a lengthscale
\begin{equation}
\ell_{\rm reg} \sim \alpha \frac{\tilde{D}}{a'}\label{eq:lreg};
\end{equation}
$\ell_{\rm reg}$ may be much larger than $\alpha$ if $\tilde{D}/a'$ is large.
Our goal is to compute 
\[
\lim\limits _{\alpha\to 0}\inf_{m}\left(\alpha I_{T}^{{\rm kin},\alpha}[\rho^\alpha,m]\right),
\]
and to minimize it over the approximating profile $\rho^\alpha$, which is the same as minimizing it over the functional $\tilde{D}[\rho]$.

Let $ m = a(\rho^\alpha) +g$. $g$ is the deviation with respect to local equilibrium, and we do not assume a priori that $g$ is of order $\alpha$.
We now split $g$ between a deterministic and a stochastic part $g=g^d+g^s$. We choose the deterministic part $g^d$ so that the rate function \eqref{eq:Kldp} vanishes when the stochastic part vanishes, ie $g^s=0$. We obtain
\begin{eqnarray}
g^d +\alpha \partial_t g^d &=& -\alpha (\partial_t a(\rho^\alpha)+\partial_x \rho^\alpha) 
\end{eqnarray}
Hence
\[
g^d =-\alpha \big(\text{Id}+\alpha \partial_t\big)^{-1}\cdot (\partial_t a(\rho^\alpha)+\partial_x \rho^\alpha).
\]
Furthermore, the constraint \eqref{eq:constraint} reads
\[
\partial_t \rho^\alpha+\partial_x\big(a(\rho^\alpha)\big) +\partial_x g^d=-\partial_x g^s
\]
ie
\begin{equation}
g^s = -g^d + \alpha \tilde{D}\partial_x \rho^\alpha;\label{eq:a-msmall}
\end{equation}
the constant which appears in the integration over $x$ is seen to vanish by considering the behavior far from the antishock: there $g=0$. 
We can now compute the rate function. First
\begin{eqnarray}
\alpha(\partial_t m +\partial_x \rho^\alpha) +m-a &=& \big(\text{Id}+\alpha \partial_t\big)\cdot g^s \nonumber \\
&=& -\big(\text{Id}+\alpha \partial_t\big)\cdot (g^d +\alpha \tilde{D}\partial_x \rho^\alpha) \nonumber \\
&=& \partial_t a(\rho^\alpha)+\partial_x \rho^\alpha +\alpha 
\tilde{D}\partial_x \rho^\alpha +\alpha^2 \partial_t (\tilde{D}\partial_x \rho^\alpha) \nonumber \\
&=& \alpha (1-a'(\rho^\alpha)^2 ) \partial_x \rho^\alpha 
+\alpha \tilde{D}\partial_x \rho^\alpha +\alpha^2 a'(\rho^\alpha)
\partial_x \tilde{D}\partial_x \rho^\alpha +\alpha^2
\partial_t \tilde{D}\partial_x \rho^\alpha \nonumber \\
&=& \alpha \left[D(\rho^\alpha)+\tilde{D}[\rho^\alpha]\right] \partial_x \rho^\alpha + \alpha^2 \left[\big(\partial_t +a'(\rho^\alpha) \partial_x\big) \tilde{D} \partial_x \rho^\alpha \right] \label{eq:gs}
\end{eqnarray}
In principle, we now have to introduce this result into \eqref{eq:Kldp}, and optimize over the functional $\tilde{D}$.
$\alpha$ always appear together with a time derivative or a spatial gradient; hence there is no small parameter left if $\ell_{\rm reg}$ is of order $\alpha$. This optimization is then a very difficult task in general, and probably not a very useful one: indeed the conditions highlighted in Sec.\ref{sec:model} for the validity of the quadratic approximation for the rate function are clearly not satisfied.
It is hard to see how this optimization could yield Jensen-Varadhan functional, but it cannot be completely ruled out.
Hence we turn in the next subsection to a regime where computations are possible, and we shall conclusively show that the limit rate function is not Jensen-Varadhan functional. 

\subsection{Computation in the $\ell_{\rm reg} \gg \alpha$ regime}
The first term in \eqref{eq:gs} is of order $\alpha \ell_{\rm reg}^{-1}$, and the second and third of order at least $\alpha^2 \ell_{\rm reg}^{-2}$ (assuming $\tilde{D}$ is of order $1$, as is $D$). It is then natural to study the regime
where $\ell_{\rm reg} \gg \alpha$, so that the first term dominates, and the second and third can be considered as perturbations. This corresponds to $a'(\rho^\alpha) \ll 1$, and this condition should be satisfied for all densities spanned by the antishock, ie $[\rho_L,\rho_R]$. Since $a'(\rho^\alpha)$ is the typical macroscopic velocity of the flow, this regime corresponds to a {\bf small Mach number}. In this regime, the left hand side of \eqref{eq:gs} is small, hence the conditions for the validity of the quadratic (i.e. Gaussian noise) approximation for the kinetic LDP  are satisfied. One may wonder if $m-a$ and 
$\alpha(\partial_t m +\partial_x \rho^\alpha)$
could be of order $1$, while their sum would be small; however, since $\partial_x \rho^\alpha$ is of order $\ell_{\rm reg}^{-1}$, from \eqref{eq:a-msmall} we obtain that $m-a$ is of order $\alpha \ell_{\rm reg}^{-1}$.

The first term in \eqref{eq:gs} is exactly the one appearing in section \ref{sec:JV}, eq.\eqref{eq:rate_inter}. Hence, at leading order in $\alpha \ell_{\rm reg}^{-1}$, one recovers Jensen-Varadhan functional: this is one of the main results of this article. Furthermore, at leading order the infimum is reached for $\tilde{D}[\rho]=D(\rho)$. Hence we may now compute the first order in $\alpha \ell_{\rm reg}^{-1}$ correction to Jensen-Varadhan functional: we only need to introduce $\tilde{D}[\rho]=D(\rho)$ into \eqref{eq:gs} and then in \eqref{eq:Kldp} \footnote{We use here the following simple remark: under appropriate regularity hypotheses, if the infimum of a functional $F_0$ is reached at $x=x_0$, then the infimum of the perturbed functional $F_0+\alpha F_1$ is $F_0(x_0)+\alpha F_1(x_0)+O(\alpha^2)$; the knowledge of $x_0$ is then sufficient to compute the first order correction to the infimum.}.   

To make the computation easier, we recall that $\rho^\alpha$ is a traveling wave regularizing the antishock: $\rho^\alpha(x,t) = U(x-v_st)$ for some profile $U$.
The velocity $v_s$ in principle depends slightly on $\alpha$, but approaches the Rankine-Hugoniot velocity in the small $\alpha$ limit. At the order of the computation, it is enough to consider $v_s$ to be fixed.
Then \eqref{eq:gs} reads
\begin{equation}
\alpha(\partial_t m +\partial_x \rho^\alpha) +m-a
= 2\alpha D(\rho^\alpha)\partial_x \rho^\alpha +\alpha^2
\big(a'(U)-v_s\big) \big(DU'\big)'.\label{eq:numerateur}
\end{equation} 
We notice that the first term is of order $\alpha \ell_{\rm reg}^{-1} \sim a'$, while the second one is of order $\alpha^2 \ell_{\rm reg}^{-2}a' \sim (a')^3$.
We also have:
\begin{equation}
(\rho^\alpha)^2 -ma(\rho^\alpha) = (\rho^\alpha)^2-(a(\rho^\alpha))^2 -\alpha a(\rho^\alpha) \tilde{D}(\rho^\alpha)\partial_x \rho^\alpha,
\label{eq:denominateur}
\end{equation}
where we have used $m=a+g$. The first two terms above in the right hand side are of order $1$, the third one is of order $\alpha \ell_{\rm reg}^{-1} \sim a'$.
Putting \eqref{eq:numerateur}-\eqref{eq:denominateur} in \eqref{eq:Kldp} and expanding with the small quantity $a'$, we see that the term of order $(a')^3$ in \eqref{eq:numerateur} does not contribute to the leading order correction
\begin{eqnarray*}
\mathcal{J}_T(\rho^A) &=& \frac{\alpha^{-1}}{4}\! \int_{0}^{T}\!\!\!\text{d}t\int_0^1\!\!\text{d}x\,\frac{\rho}{(\rho^\alpha)^2 -(a(\rho^\alpha))^2} \left[1-\alpha \frac{aD\partial_x\rho^\alpha}{(\rho^\alpha)^2 -(a(\rho^\alpha))^2}\right]^{-1}
\left[2\alpha D(\rho^\alpha)\partial_x \rho^\alpha +O\big((a')^3\big) \right]^2 \\
&=&\int_{0}^{T}\!\!\!\text{d}t\int_0^1\!\!\text{d}x\,\left[\frac{\alpha \rho}{(\rho^\alpha)^2 -(a(\rho^\alpha))^2} \big(D(\rho^\alpha)\partial_x \rho^\alpha\big)^2 +\alpha^2\frac{a\rho^\alpha}{\left[(\rho^\alpha)^2 -(a(\rho^\alpha))^2\right]^2} \big(D(\rho^\alpha)\partial_x \rho^\alpha\big)^3+O\big(\alpha^{-1}(a')^4\big) \right]
\end{eqnarray*}
The first term above yields the generalized Jensen-Varadhan functional \eqref{eq:JVLDF}, the second one provides the leading order in $a'$ correction.
Introducing the traveling wave form $\rho^\alpha(x,t) = U(x-v_st)$ into \eqref{eq:Dtilde}, one obtains
\[
\alpha D(U) U' = a(\rho_L)-a(U) +v_s (U-\rho_L).
\]
Combining this expression with a change of variable $u=U(x)$, we finally obtain, after time integration
\begin{equation}
\mathcal{J}_T(\rho^A) = I_{\rm JV}(\rho^A) + T\int_{\rho_L}^{\rho_R}\text{d}u\,\frac{4a(u)D(u)}{u \sigma^2(u)} \left[v_s(u-\rho_L)+a(\rho_L)-a(u) \right]^2+O((a')^3),
\label{eq:JV+corr}
\end{equation}
where $I_{\rm JV}$ is given by eq. \eqref{eq:JVLDF}. 
If $a'(u) \ll 1$ for any $u \in [\rho_L,\rho_R]$, then $v_s(u-\rho_L)+a(\rho_L)-a(u)$ is small; hence the correction term in \eqref{eq:JV+corr} is indeed much smaller than $I_{\rm JV}$ in this regime. At this level of approximation, $D(u)$ and $a(u)$ can actually be taken as constants. Notice $a(u)$ is not necessarily small; if it is, then the correction term in \eqref{eq:JV+corr} is smaller. The correction term in \eqref{eq:JV+corr} has been derived from the quadratic approximation of the Hamiltonian. For consistency, we should check that cubic and higher order terms in the Hamiltonian do not contribute relevant corrections to \eqref{eq:JV+corr}. To see this, notice that, as all odd order terms, the cubic coefficient in $p_m$ of $H_J$ \eqref{eq:HJ_full} is proportional to $m-a$, hence of order $a'$; hence it induces a correction in the rate function of order $a'p_m^3=O((a')^4)$, which does not affect the leading correction in \eqref{eq:JV+corr}. 

As a sanity check, it is useful to compute the rate function for an entropic shock $\rho^C$ instead of an antishock $\rho^A$. Then the first term in \eqref{eq:gs}, which is the leading order in the $\alpha \ll \ell_{\rm reg}$ regime, can be
sent to $0$ by taking $\tilde{D}[\rho]=-D(\rho)$. The leading order contribution to $\mathcal{J}_T(\rho^C)$ is then at least of order $(a')^6$. In other words we indeed find that $\mathcal{J}_T(\rho^C)$ vanishes at the order of this computation.

We have used a local expression for the rate function, taking advantage of the hypothesis that the radius of interaction is small: $R$ is at most of order $\ell$. If $R$ is of order $\ell$, this amounts to neglect terms of order $\alpha^2 \partial_{xx}\rho$: these are a priori not negligible around shocks and would enter in the computations of section \ref{sec:gen_comp}. In the $\ell_{\rm reg}\gg \alpha$ regime studied here however they would not modify \eqref{eq:JV+corr}. If $R \ll \ell$, the neglected terms are even smaller.

To summarize, we have shown on this toy example:
\begin{enumerate}
\item In general, since the correction term in \eqref{eq:JV+corr} does not vanish, the large deviations of the density are not exactly described by the Jensen-Varadhan functional \eqref{eq:JVLDF}. We also conclude that the assumption of a quadratic rate function at the kinetic level (ie Gaussian noise) is a priori not sufficient for Jensen-Varadhan functional to be valid at the hydrodynamic level.
\item However, in the regime where shocks are regularized over length scales much larger than the mean free path, which corresponds to small Mach numbers, the Jensen-Varadhan functional for the large deviation of the density is recovered. This small Mach number condition is not surprising if one wants to ensure local equilibrium for typical density profiles. We have shown here that this condition is actually also sufficient for local equilibrium to hold for non typical density profiles, at large deviation level. 
\item Finally, we have computed the first order correction to the Jensen-Varadhan functional for the cost of an antishock, \eqref{eq:JV+corr}.
\end{enumerate}

\section*{Acknowledgments}
We warmly thank C\'edric Bernardin, Freddy Bouchet and Rapha\"el Ch\'etrite for several discussions. We also thank anonymous referees for many insightful remarks. This work has been supported by the projects RETENU ANR-20-CE40-0005-01 and PERISTOCH ANR-19-CE40-0023 of the French National Research Agency (ANR).
\appendix

\section{The kinetic LDP}
\label{sec:appendixA}
We explain here the derivation of the kinetic large deviation principle \eqref{eq:LDPkinetic}-\eqref{eq:Kinetic_ratefunction} for the empirical distribution $(\hat{f}_+,\hat{f}_-)$:
\begin{align}
\hat{f}_{\pm}\left(x,t\right)=\varepsilon\sum_{n=1}^{N}\delta\left(x_n\left(t\right)-x\right),
\end{align}
where $\varepsilon^{-1} = N\ell/L$ is the number of particle in a mean free path volume.

The large deviation Hamiltonian can be computed from the following formula (see for instance \cite{FreddyBoltzmann}, and \cite{OuassimSOH,These_Ouassim} for other examples of application):
\begin{equation}
H[f_+,f_-,p_+,p_-] = \lim\limits_{\varepsilon \to 0^+} \varepsilon G\left[e^{\varepsilon^{-1}\int (p_+f_++p_-f_-)dx}\right]e^{-\varepsilon^{-1}\int (p_+f_++p_-f_-)dx},
\label{eq:gen_to_ham}
\end{equation}
where $G$ is the infinitesimal generator of the Markov process $\hat{f}_{\pm}$. $G$ acts on functionals $\phi$ of the empirical distribution $(\hat{f}_+,\hat{f}_-)$, and reads
\begin{equation}
G[\phi] = \lim\limits_{s\to 0^+} \frac{\mathbb{E}_{f_\pm}\big[\phi[f_+(s),f_-(s)]\big]- \phi[f_+,f_-]}{s},
\label{eq:gen}
\end{equation}
where $\mathbb{E}_{f_\pm}$ denotes the expectation for the process starting at $(f_+,f_-)$ at time $s=0$.

The generator can be split into two terms
\[
G_{f}=G_{T}+G_{J},
\]
where $G_{T}$ and $G_{J}$ are respectively the free transport and the jump process contributions. A Taylor expansion of $\phi\left[f\left(t\right)\right]$ at small times
allows to compute the transport part of the generator
\begin{equation}
G_{T}[\phi]=-\varepsilon^{-1}\int dx\, \left(-\partial_xf_+\frac{\delta\phi}{\delta f_+(x)}+\partial_xf_-\frac{\delta\phi}{\delta f_-(x)}\right).\label{eq:G_T}
\end{equation} 
To compute $G_{J}$, we need to evaluate the effect of jumps between $+1$ and $-1$ velocities
on the empirical distribution. The jump rates of a particle in a volume $dy$ around position $y$, respectively from velocity $+1$ to $-1$ and from velocity $-1$ to $+1$ are: 
\[
\varepsilon^{-1} f_+(y)g_-\big((\chi_R\ast \rho(y)\big)~{\rm and}~\varepsilon^{-1} f_-(y)g_+\big((\chi_R\ast \rho(y)\big).
\]
A jump from velocity $+1$ to $-1$  (resp. from $-1$ to $+1$) changes the empirical distribution as:
\[
(f_+(x),f_-(x)) \to \big(f_+(x)-\varepsilon \delta(x-y), f_-(x)+\varepsilon \delta(x-y)\big)~{\rm resp.}~(f_+(x),f_-(x)) \to \big(f_+(x)+\varepsilon \delta(x-y), f_-(x)-\varepsilon \delta(x-y)\big).
\]
Introducing this into the  expression \eqref{eq:gen} for the generator yields:
\begin{equation}
\begin{split}
G_J[\phi] = &\varepsilon^{-1}\int dy  \big\{f_+(y)g_-\big((\chi_R\ast \rho(y)\big)  \phi[f_+(\cdot)-\varepsilon \delta(\cdot-y),f_-(\cdot)+\varepsilon \delta(\cdot-y)]-\phi[f_+,f_-] \big\} \\
& + \varepsilon^{-1}\int dy  \big\{f_-(y)g_+\big((\chi_R\ast \rho(y)\big)  \phi[f_+(\cdot)+\varepsilon \delta(\cdot-y),f_-(\cdot)-\varepsilon \delta(\cdot-y)]-\phi[f_+,f_-] \big\}.
\end{split}
\label{eq:G_J}
\end{equation}
Introducing \eqref{eq:G_T} and \eqref{eq:G_J} in \eqref{eq:gen_to_ham}, we obtain the expression \eqref{eq:HT} and \eqref{eq:HJ} for the transport and jump Hamiltonians (notice $f_\pm g_\mp(\chi_R\ast \rho)= f_\pm M_\mp^R/\rho$).

\section{From the small diffusion LDP \eqref{eq:LDPalpha} to the generalized Jensen-Varadhan functional \eqref{eq:JVLDF}}
\label{sec:appendixB}

\subsection{Position of the problem}
We give in this appendix a heuristic account of the computation deriving the Jensen-Varadhan functional from the small diffusion LDP \eqref{eq:LDPalpha}. For $\rho^A$ the given antishock profile, this computation solves the problem \eqref{eq:limit}, which we recall here:
\begin{equation}
\inf_{\rho^\alpha \to \rho^A} \lim\limits_{\alpha \to 0}\alpha^\gamma \mathcal{I}_{T}^\alpha\left[\rho^\alpha \right], \label{eq:limit2}
\end{equation}
or equivalently it solves the variational problem \eqref{eq:rate_inter}.

We have to show two inequalities:
\begin{itemize}
\item The lower bound: it is not possible to approximate $\rho^A$ at a (probabilistic) cost smaller than the Jensen-Varadhan functional. In mathematical words: for any sequence $(\rho^\alpha)$ approximating $\rho^A$ in the $\alpha\to 0$ limit, 
\[
\lim\limits_{\alpha\to 0}\alpha \mathcal{I}_T^\alpha[\rho^\alpha] \geq I_{JV}[\rho^A].
\]
\item The upper bound: it is possible to approximate $\rho^A$ at a cost given by the Jensen-Varadhan functional; i.e. there exists a sequence $(\rho^\alpha)$ approximating $\rho^A$ in the $\alpha\to 0$ limit, 
\[
\lim\limits_{\alpha\to 0}\alpha \mathcal{I}_T^\alpha[\rho^\alpha] = I_{JV}[\rho^A].
\]
\end{itemize}
The computations are done here on the whole real line for simplicity, but periodic boundary conditions would not create much trouble, since all phenomena are concentrated in small neighborhoods of the shocks and antishocks. 

\subsection{The lower bound}
The starting point here is the following variational expression for $\mathcal{I}_T^\alpha$, which can be checked directly:
\begin{equation}
\alpha \mathcal{I}_T^\alpha[\rho] = \sup_\varphi \left(  \int_0^T \!\!\text{d}t\int\! \text{d}x \, \varphi \big[\partial_t \rho +\partial_x(a(\rho) -\alpha D\partial_x \rho)\big]  -\frac{\alpha}{2}  \int_0^T \!\!dt\int \!dx  \sigma(\rho) (\partial_x \varphi)^2\right).
\end{equation}
Let $\rho^\alpha$ any sequence of profiles approximating the antishock $\rho^A$.
Take $\eta(z)$ a convex real function, $q(z)$ such that $q'=a'\eta'$, and use $\varphi=\eta'(\rho^\alpha)$ in the expression above. Then
\begin{eqnarray}
\alpha \mathcal{I}_T^\alpha[\rho^\alpha] &\geq &  \int_0^T \!\!\text{d}t\int\! \text{d}x \, \eta'(\rho^\alpha) \big[\partial_t \rho^\alpha +\partial_x(a(\rho^\alpha) -\alpha D\partial_x \rho^\alpha)\big]  -\frac{\alpha}{2}  \int_0^T \!\!\text{d}t\int \!\text{d}x \,  \sigma(\rho^\alpha) (\partial_x \eta'(\rho^\alpha))^2 \\
&=&  \int_0^T \!\!\text{d}t\int\! \text{d}x \, (\partial_t \eta(\rho^\alpha) +\partial_x(q(\rho^\alpha))) + \frac{\alpha}{2} \int_0^T \!\! \text{d}t\int \!\text{d}x \, \left(2D-\sigma(\rho^\alpha) \eta''\right) \eta''(\rho^\alpha) (\partial_x \rho^\alpha)^2
\end{eqnarray}
In particular; we may choose $\eta$ such that the second term on the right hand side above vanishes:
\[
\eta''(\rho) = \frac{2D(\rho) }{\sigma(\rho)}.
\]
For this particular $\eta$ we obtain the following lower bound, valid for any sequence of profiles $\rho^\alpha$:
\begin{equation}
\alpha \mathcal{I}_T^\alpha[\rho]   \geq \int_0^T \!\!\text{d}t\int\! \text{d}x \, \big(\partial_t \eta(\rho^\alpha) +\partial_x(q(\rho^\alpha))\big).
\label{eq:borne_inf}
\end{equation}
We recall that $\rho^\alpha$ approximates the antishock $\rho^A$, hence it is close to a traveling wave $U^\alpha(x-v_s t)$, where $U^\alpha(y)$ approaches $\rho_L+(\rho_R-\rho_L) H(y)\big)$ for small $\alpha$ 
($H$ is the Heaviside step function).  
We can compute $\partial_t \eta(\rho^\alpha) +\partial_x(q(\rho^\alpha))$ in the sense of distributions. Let $g$ be a test function:
\begin{eqnarray*} 
\int_0^T \!\!\text{d}t\int\! \text{d}x \, \big(\partial_t \eta(\rho^\alpha) +\partial_x(q(\rho^\alpha))\big) g(x,t) &=& \int_0^T \!\!\text{d}t\int\! \text{d}x \,  U'(x-v_st)  \big(-v_s \eta'(U^\alpha(x-v_st) + q'(U^\alpha(x-v_st))\big) g(x,t) \\
&=& \int_0^T \!\!\text{d}t\int_{\rho_L}^{\rho_R} \text{d}u \, \big(-v_s \eta'(u) + q'(u)\big) g\big((U^\alpha)^{-1}(u)+v_s t,t\big), 
\end{eqnarray*}
where we have used the change of variable $u=U^\alpha(x-v_st)$. Since $\lim\limits_{\alpha \to 0} (U^\alpha)^{-1}(u)=0$ for all $u \in ]\rho_L,\rho_R[$:
\begin{eqnarray} 
\lim\limits_{\alpha \to 0} \int_0^T \!\!\text{d}t\int\! \text{d}x \, \big(\partial_t \eta(\rho^\alpha) +\partial_x(q(\rho^\alpha))\big) g(x,t) &=&  \int_0^T \!\!\text{d}t \, g(v_s t,t) \big[-v_s \eta'(u) + q'(u)\big]_{\rho_L}^{\rho_R} 
\end{eqnarray}
\begin{equation}
=  \int_0^T \!\!\text{d}t\int\! \text{d}x \, \delta(x-v_s t,t)  \big[-v_s \big(\eta(\rho_R)-\eta(\rho_L)\big) + q(\rho_R)-q(\rho_L)\big] g(x,t). \label{eq:app_inter}
\end{equation}
Finally, we write
\begin{eqnarray}
\eta(\rho_R)-\eta(\rho_L) &=& \int_{\rho_L}^{\rho_R} \eta'(u) \text{d}u \\
&=&\big[(u-\rho_L)\eta'(u)\big]_{\rho_L}^{\rho_R} - \int_{\rho_L}^{\rho_R} (u-\rho_L)\eta''(u) \text{d}u  \label{eq:eta}\\
q(\rho_R)-q(\rho_L) &=& \int_{\rho_L}^{\rho_R} q'(u) \text{d}u \\
&=& \int_{\rho_L}^{\rho_R} a'(u)\eta'(u) \text{d}u \\
&=&\big[(a(u)-a(\rho_L))\eta'(u)\big]_{\rho_L}^{\rho_R} - \int_{\rho_L}^{\rho_R} (a(u)-a(\rho_L))\eta''(u) \text{d}u.  \label{eq:q}
\end{eqnarray}
Putting \eqref{eq:eta} and \eqref{eq:q} into \eqref{eq:app_inter}, then into \eqref{eq:borne_inf}, we obtain
\begin{equation}
\lim\limits_{\alpha \to 0} \alpha \mathcal{I}_T^\alpha[\rho]   \geq \int_0^T \text{d}t \int_{\rho_L}^{\rho_R}  \text{d}u \, \frac{2D(u) }{\sigma(u)}\big[ v_s u -a(u) +a(\rho_L) -v_s \rho_L \big].
\label{eq:lower}
\end{equation}

\subsection{The upper bound}
The starting point for the upper bound is the expression for $\mathcal{I}_T^\alpha$ given in the main text, which we recall here:
\begin{equation}
\mathcal{I}^\alpha_{T}\left[\rho\right]= \frac12\alpha^{-2}\int_{0}^{T}\text{d}t\int_{\mathbb{R}}\text{d}x\,
\frac{1}{\sigma(\rho)}\left\{ \int^{x}\left(\partial_{t}\rho+\partial_{x}\left(a\left(\rho\right)\right)-\alpha\partial_{x}\left(D\left(\rho\right)\partial_{x}\rho\right)\right)\right\} ^{2}. \label{eq:LDPalpha_app}
\end{equation}
We now choose a particular sequence $(\rho^\alpha)$ approximating the antishock $\rho^A$. We choose $\rho^\alpha$ as the traveling wave solution of:
\begin{eqnarray}
\rho^\alpha(-\infty)=\rho_L &;&\rho^\alpha(+\infty)=\rho_R \\
\partial_t \rho^\alpha +\partial_x (a(\rho^\alpha)) &=& -\alpha \partial_x \big(D(\rho^\alpha)\partial_x \rho^\alpha\big). \label{eq:rho_app}
\end{eqnarray}
 \eqref{eq:LDPalpha_app} can then be simplified:
\begin{equation}
\mathcal{I}^\alpha_{T}\left[\rho^\alpha\right]= \int_{0}^{T}\text{d}t\int_{\mathbb{R}}\text{d}x\,
\frac{2}{\sigma(\rho^\alpha)}\left[ D\left(\rho^\alpha\right)\partial_{x}\rho^\alpha \right]^{2}. \label{eq:LDPalpha_app2}
\end{equation}
$\rho^\alpha$ is a traveling wave with velocity $v_s$ (the Rankine-Hugoniot velocity), hence we can write $\rho^\alpha(x,t) = U\big((x-v_st)/\alpha \big)$ for some profile $U$ interpolating between $\rho_L$ and $\rho_R$.  
$U$ satisfies the equation 
\[
-v_S U' +a' U' = -\big(D(U) U'\big)'.
\]
Integrating, one obtains $D(U)U' = v_sU - a(U) -v_s\rho_L+a(\rho_L)$, where the integration constant has been chosen so that $U'(-\infty)=U'(+\infty)=0$. Introducing this into  \eqref{eq:LDPalpha_app2}, and performing the change of variable $u=U\big((x-v_st)/\alpha \big)$:
\begin{equation}
\mathcal{I}^\alpha_{T}\left[\rho^\alpha\right]=\alpha^{-1} \int_{0}^{T}\text{d}t\int_{\rho_L}^{\rho_R}
\frac{2D(u)}{\sigma(u)}\left[ v_s u-a(u) -v_s \rho_L +a(\rho_L) \right] \text{d}u. \label{eq:LDPalpha_app_fin}
\end{equation}
 this last expression coincides with the lower bound \eqref{eq:lower} and, after integrating over time, with \eqref{eq:JVLDF} for antishocks. The expression in \eqref{eq:JVLDF} is slightly more complicated to make it also valid for shocks.



\begin{thebibliography}{99}
\bibitem{MFT}  L. Bertini, A. De Sole, D. Gabrielli, G. Jona-Lasinio and C. Landim Macroscopic Fluctuation Theory for Stationary Non-Equilibrium States
Journal of Statistical Physics volume 107, pages 635?675 (2002)

\bibitem{MFT2} L. Bertini, A. De Sole, D. Gabrielli, G. Jona-Lasinio and C. Landim, Macroscopic
fluctuation theory, Rev. Mod. Phys. 87, 593 (2015), doi:10.1103/RevModPhys.87.593.

\bibitem{Doyon19} B. Doyon and J. Myers, Fluctuations in ballistic transport from Euler hydrodynamics,
Ann. Henri Poincar\'e 21(1), 255 (2019), doi:10.1007/s00023-019-00860

\bibitem{Doyon22a} B. Doyon, G. Perfetto, T. Sasamoto and T.  Yoshimura, Ballistic macroscopic fluctuation theory. SciPost Physics, 15(4), 136 (2023).

\bibitem{Doyon22b} B. Doyon, G. Perfetto, T. Sasamoto and T.  Yoshimura, Emergence of hydrodynamic spatial long-range correlations in nonequilibrium many-body systems. Physical Review Letters, 131(2), 027101 (2023). 

\bibitem{SpohnBook} H. Spohn, Large Scale Dynamics of Interacting Particles, Springer Berlin Heidelberg,
doi:10.1007/978-3-642-84371-6 (1991).

\bibitem{J} L.H. Jensen, Large deviations of the asymmetric simple
exclusion process in one dimension. Ph.D. Thesis, Courant Institute
NYU (2000) 

\bibitem{JV} S.R.S Varadhan, Large deviations for the simple asymmetric
exclusion process. Stochastic analysis on large scale interacting
systems. Adv. Stud. Pure Math. 39, 1?27 (2004) 

\bibitem{QT} J. Quastel and L.-C. Tsai, Hydrodynamic Large Deviations of TASEP (2021). arXiv preprint arXiv:2104.04444

\bibitem{EnaudDerrida2004} C. Enaud and B. Derrida, Large deviation functional of the weakly asymmetric exclusion process. Journal of statistical physics, 114, 537-562 (2004).

\bibitem{BodineauDerrida2006} T. Bodineau and B. Derrida, Current large deviations for asymmetric exclusion pro-
cesses with open boundaries, J. Stat. Phys. 123(2), 277 (2006)

\bibitem{Bertini09} L. Bertini, C. Landim and M.  Mourragui, Dynamical large deviations for the boundary driven weakly asymmetric exclusion process. The Annals of Probability, 37(6), 2357-2403 (2009).

\bibitem{BertiniQP09} L. Bertini, D. Gabrielli and C. Landim, Strong asymmetric limit of the quasi-potential of the boundary driven weakly asymmetric exclusion process. Communications in Mathematical Physics, 289, 311-334 (2009).

\bibitem{Caflisch80} R.E. Caflisch, The fluid dynamic limit of the nonlinear Boltzmann equation. Communications on Pure and Applied Mathematics, 33(5), 651-666 (1980).

\bibitem{BGL91} C. Bardos, F. Golse and D. Levermore, Fluid dynamic limits of kinetic equations. I. Formal derivations. Journal of statistical physics, 63, 323-344 (1991).

\bibitem{Mariani10} M. Mariani Large deviations principles for stochastic
scalar conservation laws, Probab. Theory Relat. Fields (2010) 147: 607-648. 

\bibitem{BBMN} G. Bellettini, L. Bertini, M. Mariani
and M. Novaga, $\Gamma$-Entropy Cost for Scalar Conservation
Laws, Arch.Rat.Mech.Anal. 

\bibitem{These_Ouassim} O. Feliachi, 
From Particles to Fluids: A Large Deviation Theory Approach to Kinetic and Hydrodynamical Limits, PhD dissertation,  Universit\'{e} d'Orl\'{e}ans (2023).

\bibitem{DawsonGartner87} D.A. Dawson and J. G\"{a}rtner, Large deviations from the McKean-Vlasov limit for weakly interacting diffusions. Stochastics: An International Journal of Probability and Stochastic Processes, 20(4), 247-308 (1987).

\bibitem{Berg72} H.C. Berg and D.A. Brown, Chemotaxis in Escherichia coli analysed by three-dimensional tracking. nature, 239, 500-504 (1972).

\bibitem{Polin09} M. Polin, I. Tuval, K. Drescher, J.P. Gollub, and R.E. Goldstein, Chlamydomonas swims with two gears in a eukaryotic version of run-and-tumble locomotion. Science, 325(5939), 487-490 (2009).

\bibitem{Mariani_greek} G. Bellettini and M. Mariani, Variational convergence of multidimensional conservation laws. Bulletin of the Greek Mathematical Society, 57, 31-45 (2010).

\bibitem{BBC} J. Barr\'e, C. Bernardin, and R. Chetrite, Density large deviations for multidimensional stochastic hyperbolic conservation laws. Journal of Statistical Physics, 170, 466-491 (2018).

\bibitem{Natalini98} R. Natalini, A discrete kinetic approximation of entropy solutions to multidimensional scalar conservation laws. Journal of differential equations, 148(2), 292-317 (1998).

\bibitem{Aregba} D. Aregba-Driollet and R. Natalini, Discrete kinetic schemes for multidimensional systems of conservation laws. SIAM Journal on Numerical Analysis, 37(6), 1973-2004 (2000). 

\bibitem{FreddyBoltzmann} F. Bouchet, Is the Boltzmann equation reversible? A large deviation perspective on the irreversibility paradox. Journal of Statistical Physics, 181(2), 515-550 (2020).

\bibitem{Rezakhanlou} F. Rezakhanlou, Large deviations from a kinetic limit. Ann. Probab. 26(3), 1259-1340 (1998).

\bibitem{OuassimSOH} O. Feliachi, M. Besse, C. Nardini and J. Barr\'e, Fluctuating kinetic theory and fluctuating hydrodynamics of aligning active particles: the dilute limit. Journal of Statistical Mechanics: Theory and Experiment, 2022(11), 113207 (2022). 


\bibitem{Agranov21}  T. Agranov, S. Ro, Y.Kafri and V. Lecomte, Exact fluctuating hydrodynamics of active lattice gases typical fluctuations. Journal of Statistical Mechanics: Theory and Experiment, 2021(8), 083208 (2021).

\bibitem{Agranov22} T. Agranov, S. Ro, Y., Kafri and V. Lecomte, Macroscopic fluctuation theory and current fluctuations in active lattice gases. SciPost Physics, 14(3), 045 (2023).


\bibitem{Bellettini10} G. Bellettini, F. Caselli, M. Mariani, Quasi-potentials of the entropy functionals for scalar conservation laws. Journal of Functional Analysis, 258(2), 534-558 (2010).

\bibitem{Bahadoran10} C. Bahadoran (2010). A quasi-potential for conservation laws with boundary conditions. arXiv preprint arXiv:1010.3624.

\end{thebibliography}
\end{document}